\documentclass[preprint,showpacs,preprintnumbers,amsmath,amssymb,pra,superscriptaddress]{revtex4}

\usepackage[dvips]{graphicx} 
\usepackage{amsfonts}
\usepackage{amscd}
\usepackage{amsmath}    

\begin{document}

\title{Quantum key distribution with triggering parametric down conversion sources}


\author{Xiongfeng Ma}
\email{xima@physics.utoronto.ca}
\author{Hoi-Kwong Lo}
\email{hklo@comm.utoronto.ca}
\affiliation{%
Center for Quantum Information and Quantum Control,\\
Department of Electrical \& Computer Engineering and Department of Physics, \\
University of Toronto, Toronto, Ontario, Canada, M5S 1A7 \\
}%

\begin{abstract}
Parametric down-conversion (PDC) sources can be used for quantum key distribution (QKD). One can use a PDC source as a triggered single photon source. Recently, there are various practical proposals of the decoy state QKD with triggering PDC sources. In this paper, we generalize the passive decoy state idea, originally proposed by Mauerer and Silberhorn. The generalized passive decoy state idea can be applied to cases where either threshold detectors or photon number resolving detectors are used. The decoy state protocol proposed by Adachi, Yamamoto, Koashi and Imoto (AYKI) can be treated as a special case of the generalized passive decoy state method. By simulating a recent PDC experiment, we compare various practical decoy state protocols with the infinite decoy protocol and also compare the cases using threshold detectors and photon-number resolving detectors. Our simulation result shows that with the AYKI protocol, one can achieve a key generation rate that is close to the theoretical limit of infinite decoy protocol. Furthermore, our simulation result shows that a photon-number resolving detector appears to be not very useful for improving QKD performance in this case. Although our analysis is focused on the QKD with PDC sources, we emphasize that it can also be applied to other QKD setups with triggered single photon sources.
\end{abstract}

\maketitle

\section{Introduction}
Quantum key distribution (QKD) \cite{BB_84,Ekert_91} allows two legitimate parties, Alice and Bob, to create a random secret key even when the channel is accessible to an eavesdropper, Eve. The security of QKD is built on the fundamental laws of physics in contrast to existing classical public key encryption schemes that are based on unproven computational assumptions.  Bennett and Brassard proposed a best-known protocol --- BB84 \cite{BB_84}. Proving the security of QKD is a hard problem. Fortunately, this problem has been solved in the last decade, see for example, \cite{Mayers_01, LoChauQKD_99, ShorPreskill_00,Koashi_Uncer_06}. Many security proofs are based on the assumption of idealized QKD system components, such as a perfect single photon source and well-characterized detectors. In practice, inevitable device imperfections may compromise the security unless these imperfections are well investigated. Meanwhile, the security of QKD with realistic devices has been proven. See \cite{MayersYao_98,IndividualAttack_00,BLMS_00,ILM_07,KoashiPreskill_03,GLLP_04} for example. 

In the original proposal of BB84 protocol, a single photon source is used. Unfortunately, single photon sources are still not commercially available with current technology. Alternatively, a weak coherent state is widely used as a photon source. We call this implementation \emph{coherent state QKD}. Many coherent state QKD experiments have been done since the first QKD experiment \cite{BBBSS_92}. 

The coherent state QKD suffers from photon-number splitting (PNS) attacks \cite{HIGM_95,BLMS_00,LutkenhausJahma_02}. Nevertheless, it has been proven unconditionally secure by Inamori, L\"{u}tkenhaus and Mayers \cite{ILM_07}. This work is improved by Gottesman, Lo, L\"utkenhaus, and Preskill (GLLP) \cite{GLLP_04}, though the performance in terms of the achievable secure distance and the key rate is limited. 

Decoy state method \cite{Hwang_03} has been proposed to improve the performance of the coherent state QKD. The security of QKD with decoy states has been proven \cite{LoDecoy_03,MasterReport,Decoy_05}. The simulation result shows us that the coherent state QKD with decoy states is able to operate as good as QKD with perfect single photon sources in the sense that the key generation rates given by both setups depend linearly on the channel transmittance \cite{Decoy_05}. Afterwards, some practical decoy-state protocols with only one or two decoy states are proposed \cite{Practical_05}, see also \cite{HEHN_05,Wang_05,Wang2_05}. The experimental demonstrations for decoy state method have been done recently \cite{ZQMKQ_06,ZQMKQ60km_06,LosAlamosDecoy_07,Zeilinger_Decoy_07,PanDecoy_07,YSS_Decoy_07}.

The motivation of decoy states is to estimate the channel properties (e.g., transmittance and error proability) better. To do that, Alice uses extra states with different light intensities during key transmission. Then Alice and Bob can consider detection statistics from signal and decoy states separately, from which they can estimate the channel transmittance and error probability better. We call the situation when Alice actively prepare decoy states \textit{active decoy state} method to differentiate from the \emph{passive decoy state} method where Alice choose decoy and signal states by passive measurements. Details can be found in Section \ref{Post}. We note that in coherent state QKD, one can only use active decoy state method.

Other than the decoy state method, we remark that there are other approaches to enhance the performance of the coherent state QKD, such as QKD with strong reference pulses \cite{Koashi_04,TLMB_Strong06} and differential-phase-shift QKD \cite{IWY_DPS02}.

Besides a coherent state source, a parametric down-conversion (PDC) source can be used
for QKD as well. There are two ways to use a PDC source for QKD. The first is to use it as a triggered (heralded) single photon source. Alice detects one of the two modes from a PDC source as a trigger \footnote{See Section \ref{Impl} for the definition of a trigger.} and actively encodes her qubit information into another mode. We call this implementation \emph{triggering PDC QKD}. The second way is to use it as an entangled photon source for entanglement-based QKD protocols. See Ref.~\cite{EntanglementPDC_07} and references cited therein. We call this implementation \emph{entanglement PDC QKD}.

The triggering PDC QKD, similar to the coherent state QKD, suffers from PNS attacks. By applying the GLLP security proof, one can find that the optimal average photon number $\mu$ is in the same order of overall transmittance $\eta$. Then the key generation rate will be on the order of $\eta^2$. For a rigorous derivation, one can refer to Appendix \ref{Optimalmu}. Thus, the performance of the triggering PDC QKD is very limited.

Since decoy states idea can substantially enhance the performance of the coherent state
QKD, a natural question will be: ``can decoy states idea be applied to the triggering PDC
QKD?" The answer is \emph{yes}. One can apply the infinite decoy state idea
\cite{Decoy_05} to the triggering PDC QKD. Not surprisingly, with decoy states, the key
generation rate can be $O(\eta)$, which is the same as the order achieved by a single-photon source. Therefore, we expect the decoy state QKD to become a standard technique not only in the coherent state QKD, but also in QKD with triggering PDC sources. The infinite decoy state protocol requires an infinite number of decoy states to be used, which is not practical. A few practical decoy proposals for triggering PDC requiring a finite number of decoy states have been proposed \cite{MauererSilberhorn_07,AYKI_07,WWG_07,WWBK_PDC_07}.

We are interested in comparing various protocols for the triggering PDC QKD. Among the practical decoy protocols for triggering PDC QKD, we find that the one proposed by Adachi, Yamamoto, Koashi and Imoto (AYKI) \cite{AYKI_07} is simple to implement. The AYKI protocol is conceptually similar to the one-decoy-state scheme \cite{Practical_05}. In the AYKI protocol, Alice and Bob only need to consider the statistics of triggered and non-triggered detection events \footnote{In a non-triggered detection event, Bob gets a detection but Alice doesn't get a trigger.} separately, instead of preparing new signals for decoy states. We emphasize that the AYKI protocol is easy to implement since there is no need for a hardware change.



Other decoy state proposals for the triggering PDC QKD require hardware modifications. For example, the one proposed by Mauerer and Silberhorn \cite{MauererSilberhorn_07} requires photon-number resolving detectors, and the one proposed by Wang, Wang and Guo \cite{WWG_07} requires Alice's pumping the laser source at various intensities.

We generalize the passive decoy state idea proposed by Mauerer and Silberhorn \cite{MauererSilberhorn_07}. The main idea is that, Bob can group his detection events according to the public announcement of Alice's detection events. For example, when Alice uses a threshold detector, Bob can group his detection results according to whether Alice gets a detection or not.  The generalized passive decoy state idea can be applied to both cases of using threshold detectors and photon-number resolving detectors. The AYKI protocol can be treated as a special case of the generalized passive decoy state protocol. 



By simulating a recent PDC experiment \cite{PDC144_07}, we compare one case with a perfect photon-number resolving detector and four cases with threshold detectors: no decoy, infinite decoy, weak decoy and AYKI.
Our simulation result shows that in a large parameter regime, the performance of AYKI protocol is close to that of the infinite decoy protocol and thus there is not much room left for improvement after the AYKI protocol has been implemented. Also, the QKD performance of the case with the infinite decoy protocol using threshold detectors is close to the case using a perfect photon-number resolving detector. Thus, a photon-number resolving detector appears to be not very useful for triggering PDC QKD.

We emphasize that one advantage of passive decoy state method is that by passively choosing decoy and signal states, the possibility that Eve can distinguish decoy and signal states is reduced. On the other hand, in active (regular) decoy state experiments, it is more difficult to verify the assumption that Eve cannot distinguish decoy and signal states.

We note that the passive decoy state idea can be combined to the active decoy state idea. In Ref.~\cite{WWBK_PDC_07}, the authors gave a special case of combining passive and active decoy state ideas. Note that for coherent state QKD, one can only use active decoy state methods.

Although our analysis is focussed on the QKD with PDC source, we emphasize that it can also
be applied to QKD setups with other triggered single photon sources.



In Section \ref{Impl}, we will review the experiment setup of the triggering PDC QKD. In Section \ref{Model}, we give a model for the triggering PDC QKD. In Section \ref{Post}, we will study various post-processing schemes for the triggering PDC QKD. In Section \ref{Simulation}, we will compare various schemes of the triggering PDC QKD: non-decoy+threshold detectors, infinite decoy+threshold detectors, AYKI and the case with a perfect photon-number resolving detector, by simulating a real PDC experiment. In Appendix \ref{Optimalmu}, we consider the optimal PDC source intensities for the triggering PDC QKD.



\section{Experiment setup} \label{Impl}
\begin{figure}[hbt]
\centering \resizebox{12cm}{!}{\includegraphics{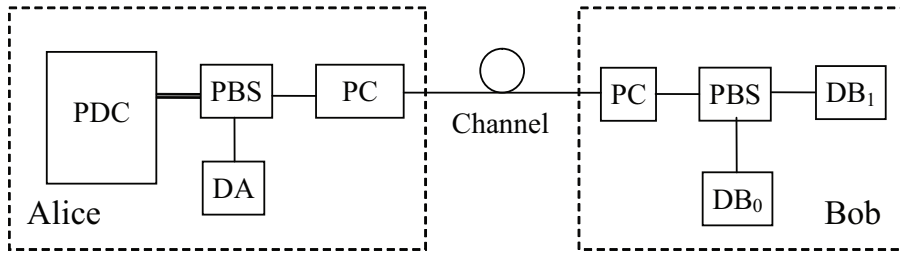}} \caption{A schematic diagram for the
triggering PDC QKD. Alice collects photon pairs emitted from a PDC source and uses a polarized beam splitter (PBS) to separate two polarization modes. She detects one of the two modes with her
detector (DA) as a trigger, modulates the polarization of the other mode by a polarization
controller (PC) and sends it to Bob. On Bob's side, he chooses his basis by a PC and performs a
measurement by his detectors (DB$_0$ and DB$_1$).} \label{Fig:PDCtrig}
\end{figure}

In triggering PDC QKD, a PDC source is used as a triggered single photon source \footnote{Sometimes it is called heralded single photon source.}. 
The schematic diagram is shown in FIG.~\ref{Fig:PDCtrig}.

As shown in FIG.~\ref{Fig:PDCtrig}, a PDC source generates two modes of photons, which can be separated by a polarized beam splitter (PBS). One mode goes to Alice's own detector (DA in FIG.~\ref{Fig:PDCtrig}) as the triggering signal and the other mode is used as a triggered single photon state for the QKD.
When Alice's detector (DA) clicks, we call it a \emph{trigger}. We divide the detection
events on Bob's side into two groups depending on whether Alice gets a trigger or not: triggering
detection events and non-triggering detection events.

Note that Alice can use either a threshold detector or a photon-number resolving detector (DA in FIG.~\ref{Fig:PDCtrig}). She only needs to know the number of photons in the trigger mode. So only one detector is sufficient on Alice's side. Due to the high channel losses, without Eve's interference, Bob is highly likely to receive a vacuum or single photon state. Thus it is sufficient for Bob to use threshold detectors. Threshold single photon detectors can only tell whether there a click or not, but not the photon numbers. Bob needs to identify polarizations of incoming photons. Here we assume Alice encodes qubit information in photon polarizations.

In real experiments, there are two types of PDC sources. In triggering PDC QKD, both of these two types can be used. Here we assume Alice uses type-II PDC source. The Hamiltonian of the type-II PDC process in the triggering setup shown in FIG.~\ref{Fig:PDCtrig} can be
written as \cite{WallsMilburn_94}
\begin{equation}\label{Impl:HamiltonianTrig}
\begin{aligned}
H &= i\chi a^\dagger b^\dagger+h.c. \\
\end{aligned}
\end{equation}
where $h.c.$ means Hermitian conjugate and $\chi$ is a coupling constant which depends on the
crystal nonlinearity and the amplitude of the pump beam. The operators $a^\dagger$, $b^\dagger$ and $a$, $b$ are the creation and annihilation operators of two modes with different polarizations.

The state coming from a triggering PDC source, with a Hamiltonian of
Eq.~\eqref{Impl:HamiltonianTrig}, can be written as \cite{WallsMilburn_94}
\begin{equation}\label{Impl:PDCstateL}
\begin{aligned}
|\Psi\rangle=(\cosh\chi)^{-1}\sum_{n=0}^{\infty}(\tanh\chi)^n|n,n\rangle.
\end{aligned}
\end{equation}
Here we assume the state is single-mode. The expected photon pair number is given by $\mu=\sinh^2\chi$.  
The probability to get an $n$-photon-pair is
\begin{equation}\label{Impl:PnL}
\begin{aligned}
P(n)=\frac{\mu^n}{(1+\mu)^{n+1}}.
\end{aligned}
\end{equation}
Here we assume that the PDC source always sends out photon pairs. That is, the photon number of mode $a$ and $b$ are always the same.

There is a nonzero probability for the PDC source to emit more than one photon pairs in
one pulse. Thus, Alice may send out multi photon states after she encodes basis and key
information by her polarization controller (PC). This is the reason why the triggering
PDC QKD suffers from PNS attacks.

Let us compare triggering PDC QKD and entanglement PDC QKD implementations. For the setup of entanglement PDC QKD, one can refer to Ref.~\cite{EntanglementPDC_07}. In the triggering PDC QKD Alice actively encodes the key information, while in the entanglement PDC QKD Alice measures the polarization of one mode of PDC source directly. The advantage of the triggering PDC QKD here is that it does not rely on the polarization correlations between two modes of the PDC source. It only requires photon-pair generation of the source, which means entanglement between photon pairs are not important for triggering PDC QKD. However, in entanglement PDC QKD implementation, the entanglement between two modes has to be well maintained for QKD transmission. We notice that maintaining entanglement in real experiments is a highly non-trivial task \footnote{We thank A.~M.~Steinberg for enlightening discussions.}.

\section{Model} \label{Model}
L\"utkenhaus has already studied the model of triggering PDC QKD \cite{IndividualAttack_00} with threshold detectors. His model is similar to the one of the coherent state QKD, except for a different photon number distribution. For the model of the coherent state QKD, one can refer to Ref.~\cite{IndividualAttack_00, Practical_05}. For the model of entanglement PDC QKD, one can refer to Ref.~\cite{EntanglementPDC_07}.

\subsection{Photon number channel model}
Here we will use the photon number channel model \cite{Decoy_05}: Alice and Bob have infinite number of channels. For channel $i$, Alice uses $i$-photon states (Fock states) to carry the qubit information, with $i=0,1,2\cdots$. The $i$th channel corresponds to the case when Alice's photon source emits an $i$-photon state. Thus, the probability for Alice to use the $i$th channel is determined by the photon source. For example, in coherent state QKD, the probability for Alice using the different channels follows a Poisson distribution. For the details of the photon number channel model, one can refer to Ref.~\cite{Practical_05}.

We define the yield $Y_{i}$ to be the probability for Bob to get a detection event conditioned on Alice using the $i$th channel. As discussed in Section \ref{Impl}, we assume that Bob uses a threshold detector. The yield is given by
\begin{equation}\label{Model:Yi}
\begin{aligned}
Y_{i} &= 1-(1-Y_{0B})(1-\eta)^i, \\
\end{aligned}
\end{equation}
where $Y_{0B}$ is the background count rate of Bob's detection system, and $\eta$ is the overall detection probability for Bob, which takes into account the channel transmission efficiency, the coupling efficiency, the detector efficiency and the other losses in Bob's box.

The error rate when Alice uses $i$th channel is given by
\begin{equation}\label{Model:enT}
\begin{aligned}
e_iY_i 
&= e_dY_i+(e_0-e_d)Y_{0B} \\
\end{aligned}
\end{equation}
where $e_0=1/2$ is the error rate of background counts, $e_d$ is the intrinsic detector error rate on Bob's side (e.g., due to misalignment) and $Y_{i}$ is given by Eq.~\eqref{Model:Yi}. Here, we neglect the case where both background counts and true signal click since $\eta$ and $Y_{0B}$ are small. We remark that Eqs.~\eqref{Model:Yi} and \eqref{Model:enT} are true for both triggered and non-triggered cases. 

\subsection{On Alice's side} \label{Aliceside}
In the triggering PDC QKD, Alice may use either a threshold detector or a photon-number resolving detector. Define an \emph{N-photon-resolving} detector to be a detector that can tell 0, 1, $\cdots$, $N$ photons of incoming signal. For a threshold detector, we have $N=1$, which can only tell there are photons presenting or not. Given an incoming $i$-photon state, the probability for Alice's detector to indicate a $j$-photon state is $\eta_{j\mid i}$, with $\sum_{j=0}^{j=N}\eta_{j\mid i}=1$ for all $i=0,1,\cdots$. In general, $\eta_{j\mid i}$'s are real numbers in [0,1]. We define a $j$-photon trigger for the case that Alice's detector indicates a $j$-photon state.

For a triggered PDC photon source, as given in Eq.~\eqref{Impl:PDCstateL}, the probability for Alice's detector to indicate a $j$-photon detection is
\begin{equation}\label{Model:AliceProbj}
\begin{aligned}
P_{Aj} &= \sum_{i=0}^{\infty}\frac{\mu^i}{(1+\mu)^{i+1}}\eta_{j\mid i}. \\
\end{aligned}
\end{equation}
With the assumption that the PDC source always emits photon pairs, the probability (gain) for Alice getting a $j$-photon detection and Bob getting a detection is
\begin{equation}\label{Model:Gainj}
\begin{aligned}
Q_{\mu,j} &= \sum_{i=0}^{\infty}Q_{i,j} \\
&= \sum_{i=0}^{\infty}\frac{\mu^i}{(1+\mu)^{i+1}}\eta_{j\mid i}Y_{i}, \\
\end{aligned}
\end{equation}
where the yield $Y_{i}$ is given in Eq.~\eqref{Model:Yi}. 
%
%
%
The quantum bit error rate (QBER) conditioned on Alice's $j$-photon detection, similar to Eq.~\eqref{Model:Gainj}, is given by
\begin{equation}\label{Model:QBERj}
\begin{aligned}
E_{\mu,j}Q_{\mu,j} &= \sum_{i=0}^{\infty}Q_{i,j}e_i \\
&= \sum_{i=0}^{\infty}\frac{\mu^i}{(1+\mu)^{i+1}}\eta_{j\mid i}Y_{i}e_i. \\
\end{aligned}
\end{equation}
where the error rate $e_{i}$ is given in Eq.~\eqref{Model:enT}.

One observation is that in the triggering PDC QKD setup, shown in FIG.~\ref{Fig:PDCtrig}, the quantities $Y_i$ and $e_i$ are independent of Alice's measurement outcome $j$. This is based on the single-mode PDC source assumption described in Eq.~\eqref{Impl:HamiltonianTrig} in Section \ref{Impl}. 
Therefore, in Section \ref{Post}, we can apply the decoy state idea.

\subsection{Threshold detector}
Here we will discuss a special case that Alice uses a threshold detector. That is,
\begin{equation}\label{Model:thresh}
\begin{aligned}
\eta_{0\mid i} &= (1-Y_{0A})(1-\eta_A)^i \\
& \simeq (1-\eta_A)^i \\
\eta_{1\mid i} &= 1-\eta_{0\mid i} \\
\eta_{j\mid i} &= 0, \text{\space\space\space\space\space\space} \forall j\ge2, \\
\end{aligned}
\end{equation}
where $Y_{0A}$ and $\eta_A$ are the background count rate and the detector efficiency on Alice's side. The approximation is due to the fact that normally we have $\eta_A\gg Y_{0A}$. That is, we neglect the background contributions on Alice's side.


According to Eqs.~\eqref{Model:Gainj} and \eqref{Model:QBERj}, without Eve's interference, the gains and QBER's of triggered ($j=1$) and non-triggered ($j=0$) detections are given by
\begin{equation}\label{Model:QEThreshold}
\begin{aligned}
Q_{\mu,0} 
&= \frac{1}{1+\eta_A\mu}-\frac{1-Y_{0B}}{1+(\eta_A+\eta-\eta_A\eta)\mu} \\
Q_{\mu,1} 
&= 1-\frac{1}{1+\eta_A\mu}-\frac{1-Y_{0B}}{1+\eta\mu}+\frac{1-Y_{0B}}{1+(\eta_A+\eta-\eta_A\eta)\mu} \\
E_{\mu,0}Q_{\mu,0} &= e_dQ_{\mu|0} + \frac{(e_0-e_d)Y_{0B}}{1+\eta_A\mu} \\
E_{\mu,1}Q_{\mu,1} &= e_dQ_{\mu|1} + \frac{(e_0-e_d)\eta_A\mu Y_{0B}}{1+\eta_A\mu}. \\
\end{aligned}
\end{equation}
Without Eve's interference, the gains and error rates of the single photon state in two detections are given by
\begin{equation}\label{Model:qe1Threshold}
\begin{aligned}
Q_{1,0} &= \frac{\mu(1-\eta_A)}{(1+\mu)^2}Y_{1} \\
Q_{1,1} &= \frac{\mu\eta_A}{(1+\mu)^2}Y_{1} \\
e_1Y_1 
&= e_dY_1+(e_0-e_d)Y_{0B} \\
\end{aligned}
\end{equation}
where $Y_1$ and $e_1$ are given in Eq.~\eqref{Model:Yi} and \eqref{Model:enT}, respectively.

\subsection{Perfect photon-number resolving detector} \label{PrfctPNRD}
Here we will discuss the case that Alice uses a perfect photon-number resolving detector, which can faithfully tell the number of photons in the incoming signal. That is, $\eta_{j\mid i}=\delta_{ij}$. Thus the gains and QBER's are given by, from Eqs.~\eqref{Model:Gainj} and \eqref{Model:QBERj},
\begin{equation}\label{Model:PrfctGQj}
\begin{aligned}
Q_{\mu,i} = Q_{i,i} 
&= \frac{\mu^i}{(1+\mu)^{i+1}}Y_{i} \\
E_{\mu,i}Q_{\mu,i} = e_iQ_{i,i} &= \frac{\mu^i}{(1+\mu)^{i+1}}e_iY_{i}, \\
\end{aligned}
\end{equation}
from where one can directly infer the gains and error rates of $i$-photon state, $Q_{i,j}=Q_{i,i}\delta_{i,j}$.

\section{Post-processing} \label{Post}
In the following discussion, we will focus on BB84 protocol \cite{BB_84}. Due to PNS attacks \cite{HIGM_95,BLMS_00,LutkenhausJahma_02}, only vacuum states and single photon states are secure for BB84 protocol, which may not be true for other protocols, such as SARG04 \cite{SARG_04}. 

Similar to the coherent state QKD, we can apply GLLP \cite{GLLP_04} security analysis to the triggering PDC QKD. First, Alice and Bob perform error correction, after which they will share an identical key. Then, they perform privacy amplification to different types of qubits separately. Since here we assume only vacuum states and single photon states are secure for BB84 protocol, the key generation rate is given by \cite{Decoy_05,Vacuum_05,Koashi_NewModel_06}
\begin{equation} \label{Post:KeyTrig}
R \geq q \{-f(E_{\mu})Q_{\mu}H_2(E_{\mu})+Q_1[1-H_2(e_1)]+Q_0\},
\end{equation}
where $q$ is the basis reconciliation factor (1/2 for the BB84 protocol due to the
fact that half of the time Alice and Bob disagree with the bases, and if one uses
the efficient BB84 protocol \cite{EffBB84_05}, $q\approx1$), the subscript $\mu$
denotes for the expected photon pair number, $Q_{\mu}$ and $E_{\mu}$ are the overall
gain and QBER, $Q_1$ and $e_1$ are the gain and error rate of single photon states,
$Q_0$ is the gain of vacuum states, $f(x)$ is the bi-direction error correction
efficiency (see, for example, \cite{BrassardSalvail_93}) as a function of the error
rate (normally $f(x)\ge1$ with the Shannon limit $f(x)=1$) and $H_2(x)=-x\log_2(x)-(1-x)\log_2(1-x)$ is the binary entropy function.

All the classical data measured can be grouped according to Alice's detection
events, $j=0,1,\cdots,N$. Then we can apply the GLLP idea \cite{GLLP_04,TwoWay_06}
to each group. The final key generation rate will be given by summing over contributions
from all groups,
\begin{equation} \label{Post:KeySum}
R = \sum_{j=0}^{N} R_j.
\end{equation}
In each case $j$, one can apply Eq.\eqref{Post:KeyTrig},
\begin{equation} \label{Post:Keyj}
R_j \geq q \{-f(E_{\mu,j})Q_{\mu,j}H_2(E_{\mu,j})+Q_{1,j}[1-H_2(e_{1})]+Q_{0,j}\},
\end{equation}
where $Q_{0,j}$ and $Q_{1,j}$ are the first and second terms in the right hand side of Eq.~\eqref{Model:Gainj}. Here the error rate of single photon state $e_1$ is independent of $j$, see the observation in the end of Section \ref{Aliceside}. We note that the key generation rate from all $j$-photon trigger detections should be non-negative. If any of them contributes a negative key generation rate, we should assign $0$ to it. In this case, Alice and Bob can just discard that type of detections. Based on this observation, we can further simplify Eq.~\eqref{Post:KeySum}. Given Alice detects more than one photons, the probability of emitting single photon state in Bob's arm is small \footnote{In Section \ref{Impl}, we assume that Alice's PDC source always sends out photon pairs. Given that Alice detects more than one photons on the triggering arm, a single photon state presents on the other arm only when there is a dark count in Alice's detector. Normally, we can assume that the detector efficiency is much higher than the dark count probability on Alice's side. Thus, we neglect probability of a single photon state with a multi photon trigger.}. As we mentioned in the beginning of this section, only single photon state can contribute positively to the final key rate.
Thus we can focus on the case $j=0,1$.
\begin{equation} \label{Post:Key01}
R = R_0+R_1,
\end{equation}
where $R_0$ and $R_1$ are given in Eq.~\eqref{Post:Keyj}. Again, both $R_0$ and
$R_1$ should be non-negative, otherwise should be assigned 0.

In Eq.~\eqref{Post:Keyj}, the gain $Q_{\mu,j}$ and the QBER $E_{\mu,j}$, given in Eqs.~\eqref{Model:Gainj} and \eqref{Model:QBERj}, can be measured or tested from QKD experiments directly. In this section, we will discuss various ways to estimate $Q_{0,j}$, $Q_{1,j}$, and $e_{1}$. We assume that the PDC photon source  and detector characteristics are fixed and known to Alice. That is, $\mu$, the photon number distribution in Eq.~\eqref{Impl:PnL} and $\eta_A$ are fixed and known.



\subsection{Non-decoy states with threshold detectors}
Here we assume that Alice uses a threshold detector. Thus, Alice only has two measurement outcomes, $j=0,1$. One simple way to estimate $Q_{0,j}$, $Q_{1,j}$, and $e_{1}$ is by assuming that all losses and errors come from the single photon states. This is because Eve can in principle perform PNS attacks on the multi-photon states.
The gain and error rate of the single photon states in triggered ($j=1$) and non-triggered ($j=0$) detections can be bounded by
\begin{equation}\label{Post:Q1e1nondecoy}
\begin{aligned}
Q_{1,0} &\ge Q_{\mu,0}-\sum_{i=2}^{\infty}\frac{\mu^i}{(1+\mu)^{i+1}}\eta_{0|i} \\
      &= Q_{\mu,0}-\frac{(1-\eta_A)^2\mu^2}{(1+\eta_A\mu)(1+\mu)^2} \\
Q_{1,1} &\ge Q_{\mu,1}-\frac{\eta_A(2-\eta_A+\mu)\mu^2}{(1+\eta_A\mu)(1+\mu)^2} \\
e_{1,0} &\ge \frac{E_{\mu,0}Q_{\mu,0}}{Q_{1,0}} \\
e_{1,1} &\ge \frac{E_{\mu,1}Q_{\mu,1}}{Q_{1,1}} \\
%
\end{aligned}
\end{equation}
where $\eta_A$ is the efficiency of Alice's detector. The gain $Q_\mu$ and the QBER $E_{\mu}$, given in Eqs.~\eqref{Model:Gainj} and \eqref{Model:QBERj}, can be measured or tested from QKD experiments directly. In the following simulations, we will use Eqs.~\eqref{Model:QEThreshold}. Since we assume all errors come from the single photon states, one should take the lower bound of the vacuum contribution to be $Q_{0,j}=0$.


\subsection{Infinite active decoy state with threshold detectors}
To do privacy amplification, Alice and Bob need to bound $Q_{0,j}$, $Q_{1,j}$, and $e_{1}$ for Eq.~\eqref{Post:Keyj}. 
From Eq.~\eqref{Model:Gainj}, we know that to bound $Q_{0,j}$ and $Q_{1,j}$, Alice and Bob need to estimate $Y_1$.

Decoy state method provide a good way to estimate $Y_1$ and $e_1$ \cite{Hwang_03,Decoy_05}. The essential idea is that instead of considering each linear equation of $Y_1$ and $e_1$ in the form of Eqs.~\eqref{Model:Gainj} and \eqref{Model:QBERj} separately, Alice and Bob consider all the linear equations simultaneously.

Let us imagine that Alice and Bob obtain an infinite number of linear equations in the form of Eqs.~\eqref{Model:Gainj} and \eqref{Model:QBERj}, e.g., they use an infinite number of intensities $\mu$. In principle, Alice and Bob can solve the equations to get $Y_1$ and $e_1$ accurately. Mathematically, the problem is solvable. The intuition is that the contributions from higher order terms of $Y_i$ and $e_i$ decrease exponentially in Eqs.~\eqref{Model:Gainj} and \eqref{Model:QBERj}. For the case coherent state QKD, one or two decoy states are proven to be sufficient \cite{Practical_05}. Shortly, we will see that one decoy state is sufficient for triggering PDC QKD.

We remark that the key underlying assumption of the decoy state method is \cite{Decoy_05}
\begin{equation}\label{Passive:DecoyAss}
\begin{aligned}
Y_i(decoy)=Y_i(signal) \\
e_i(decoy)=e_i(signal).
\end{aligned}
\end{equation}
In another word, Eve sets the same values of $Y_i$ and $e_i$ for decoy and signal states. This can be guaranteed by the assumption that Eve cannot distinguish decoy and signal states.

In Appendix \ref{Optimalmu}, we will show that the optimal $\mu$ for the infinite decoy state case is in the order of 1, $\mu=O(1)$, which yields final key rate $R=O(\eta)$. On the other hand, the optimal $\mu$ for non-decoy case is $\mu=O(\eta)$, which yields final key rate $R=O(\eta^2)$. Therefore, we expect the decoy state QKD to become a standard technique not only in the coherent state QKD, but also in QKD with triggering PDC sources.


There are various ways to apply the decoy state idea to the triggering PDC QKD \cite{MauererSilberhorn_07,AYKI_07,WWG_07}. Here we consider the upper bound (infinite decoy state case) of all possible decoy protocols of triggering PDC QKD with threshold detectors: triggering PDC+infinite decoy method \cite{Decoy_05}. In the infinite decoy state method, Alice and Bob perform infinite number of decoy states by choosing different intensities of the PDC source, $\mu$. Then they can solve the linear equations in the form of Eqs.~\eqref{Model:Gainj} and \eqref{Model:QBERj} to estimate $Y_1$ and $e_1$ accurately. So they can calculate each $Q_{0,j}$, $Q_{1,j}$, and $e_{1}$ accurately. In the simulation, we will use Eqs.~\eqref{Model:QEThreshold} and \eqref{Model:qe1Threshold} directly.


\subsection{Weak active decoy state with threshold detectors}
Here we assume that Alice and Bob use threshold detectors and focus on triggered detection events. Alice uses another intensity $\nu$, say by attenuating pumping laser, for the weak decoy state. Wang, Wang and Guo have proposed a practical decoy method for triggering PDC QKD \cite{WWG_07}, which is essentially applying vacuum+weak decoy state method \cite{Practical_05}. Note that for triggered detection events, the vacuum contribution can be negligible since $\eta_A\gg Y_{0A}$. Thus there is no need to estimate the vacuum contribution here. So Alice and Bob only need to perform weak decoy state instead of vacuum+weak decoy states. In this case, only one weak decoy state is sufficient.

Bounds of $Y_1$ and $e_1$ are given by $\mu^2(1+\nu)^3\times Q_{\nu,1}-\nu^2(1+\mu)^3\times Q_{\mu,1}$ in Eq.~\eqref{Model:Gainj} and Eq.~\eqref{Model:QBERj}
\begin{equation}\label{Post:ActiveY1e1}
\begin{aligned}
Y_1 &\ge \frac{1}{\eta_A(\mu-\nu)}[\frac{\mu}{\nu}(1+\nu)^3Q_{\nu|1}-\frac{\nu}{\mu}(1+\mu)^3Q_{\mu|1}] \\
e_1 &\le \min\{\frac{(1+\mu)^2}{\mu}\frac{E_{\mu,1}Q_{\mu,1}}{\eta_AY_{1}}, \frac{(1+\nu)^2}{\nu}\frac{E_{\nu,1}Q_{\nu,1}}{\eta_AY_{1}}\} \\
\end{aligned}
\end{equation}
where $\nu$ is the expected photon pair number of the weak decoy state and $\eta_A$
is the efficiency of Alice's threshold detector. 

It is not hard to show that when $\nu\rightarrow0$, Eq.~\eqref{Post:ActiveY1e1} approaches the infinite case, Eqs.~\eqref{Model:QEThreshold} and \eqref{Model:qe1Threshold}, described in the previous subsection.

\subsection{Passive decoy state} \label{Sub:Passive}
Recently, Mauerer and Silberhorn proposed a passive decoy state scheme, in which photon-number resolving detectors are required \cite{MauererSilberhorn_07}. Let us recap the heuristic idea of the original passive decoy state scheme briefly here. As discussed in the Section \ref{Model}, Alice and Bob eventually get different detection events grouped by triggers on Alice's side. The key idea proposed by Mauerer and Silberhorn is that Alice and Bob manually combine the $\{j\}$-trigger detection events to get the decoy states with different photon number statistics and then follows regular decoy state scheme.

Here we want to point out that the ``combination" step is unnecessary. In general, each detection event group with $j$-trigger has a different photon number statistics on photon source arm. Thus, what Alice and Bob need to do is treating all $\{j\}$-trigger detection events statistics separately. Furthermore, photon-number resolving detectors are not necessary in passive decoys state scheme. Our new generalized passive decoy state scheme is as follows.
\begin{enumerate}
\item
Alice uses a PDC source as her triggered photon source. She detects one of the modes from her PDC source as trigger and encode key information into another mode. Due to the detector Alice uses, she will get different trigger events: $j=0,1,\cdots$. When she uses a threshold detector, she will only get $j=0,1$.

\item
As usual BB84 protocol, Bob measures signals in two different bases. Alice and Bob perform basis reconciliation.

\item
Alice announces her trigger detection results for each pulse: $j$. Bob group his detection events by the information $j$. For each $j$, they calculate the gain $Q_{\mu,j}$ and test the QBER $E_{\mu,j}$.

Mathematically, they will obtain a set of linear equations in the form of Eqs.~\eqref{Model:Gainj} and \eqref{Model:QBERj}. Notice that the setup parameters, $\mu$ and $\eta_{j\mid i}$'s, are known to Alice and Bob. Thus, they can estimate $Y_1$ and $e_1$ by considering Eqs.~\eqref{Model:Gainj} and \eqref{Model:QBERj}. 

\item
Apply post-processing according to Eq.~\eqref{Post:Key01}.
\end{enumerate}

We remark that the scheme is called \emph{passive} because Alice does not actively select decoy states. Instead, she determines the decoy states by measuring the trigger mode. Later, we will show that this is one advantage of using triggering PDC source for QKD. Actually, in this case,  there is no strict definitions of decoy states and signal states. In the original decoy state method \cite{Practical_05}, decoy states are only used to estimate $Y_1$ and $e_1$ and the key is always generated from signal states \footnote{In the coherent state QKD, there is an optimal $\mu$ for a setup. To maximize the final key rate, Alice and Bob should publicly compare all detection results from decoy states.}.  In triggering PDC QKD case, both the triggered $j=0$ and non-triggered $j=1$ detection events may have positive contribution to the final key generation.

\subsection{Passive decoy state with threshold detectors} \label{AYKI}
Here we will review the decoy protocol proposed by Adachi, Yamamoto, Koashi and Imoto \cite{AYKI_07} as a special case of the passive decoy state protocol. The AYKI protocol is interesting in practice since it doesn't involve any hardware change to implement decoy state.

Both Alice and Bob use threshold detectors, thus they have two types of detection events, triggered ($j=1$) and non-triggered ($j=0$). Secure keys can be generated from both types of detection events. Following the passive decoy state method procedure described in the previous subsection, Alice and Bob can estimate $Y_1$ and $e_1$ by considering the statistics of triggered and non-triggered detection events together. This is conceptually similar to one decoy state idea \cite{Practical_05}.

By solving two linear equations of Eq.~\eqref{Model:Gainj} with $j=0,1$,
$[1-(1-\eta_A)^2]\times Q_{\mu,0}-(1-\eta_A)^2\times Q_{\mu,1}$, one can get
\begin{equation}\label{Post:Y1T}
\begin{aligned}
Y_1 &\ge Y_1^L \equiv \frac{(1+\mu)^2}{\mu}[\frac{2-\eta_A}{1-\eta_A}(Q_{\mu,0} -Q_{0,0})-\frac{1-\eta_A}{\eta_A}Q_{\mu,1}] \\
\end{aligned}
\end{equation}
where $Q_{0,0}$ is the vacuum state contribution in non-triggered detection events. One need to minimize the key rate of Eq.~\eqref{Post:Key01} for $Q_{0,0}$ with the constraint of Eq.~\eqref{Model:QBERj}. We note that this result is essentially the Eq.~(14) given in Ref.~\cite{AYKI_07}. We can see that when $\eta_A$ is close to 1 or $\mu$ is small, after neglecting $Q_{\mu,0}$ (background counts), the lower bound $Y_1^L$ is tight (approaches the real value of $Y_1$, see Eq.~\eqref{Model:Yi}),
\begin{equation}\label{Post:Y1Tlim}
\begin{aligned}
\lim_{\eta_A\rightarrow1}Y_1^L = \lim_{\mu\rightarrow0}Y_1^L &= \eta. \\
\end{aligned}
\end{equation}


By neglecting the vacuum state contribution for triggered detection events, $Q_{0,1}=0$, $e_1$ can be simply estimated by
\begin{equation}\label{Post:e1U}
\begin{aligned}
e_1 \le 
\frac{E_{\mu,1}Q_{\mu,1}}{Q_{1,1}}.
%
%
\end{aligned}
\end{equation}

To get the lower bound of $Y_1$ in Eq.~\eqref{Post:Y1T}, one has to estimate the background contribution $Q_{0,0}$ as well. One simple bound of $Q_{0,0}$ is $0\le Q_{0,0}e_0\le E_{\mu,0}Q_{\mu,0}$ from Eq.~\eqref{Model:QBERj}, where $e_0=1/2$.

We note that the key rate calculated by substituting Eqs.~\eqref{Post:Y1T} and \eqref{Post:e1U} into Eq.~\eqref{Post:Key01} is not optimal. To get a tighter key rate bound, one can numerically lower bound Eq.~\eqref{Post:Key01} directly given the measurement results, Eq.~\eqref{Model:qe1Threshold}.

\subsection{With a perfect photon-number resolving detector}
Here we discuss a special case that Alice uses a perfect photon-number resolving detector, discussed in Section \ref{PrfctPNRD}.
Now that Alice knows the exact photon number of the source, Alice and Bob only need to focus the post-processing on the single photon state detection events. In this case, the BB84 protocol is implemented by single photon states only. Thus, they can directly apply Shor and Preskill's formula \cite{ShorPreskill_00,EntanglementPDC_07}
\begin{equation} \label{Post:KeyTrig}
R \geq qQ_1 [1-f(e_1)H_2(e_1)-H_2(e_1)].
\end{equation}
Later from the simulation, shown in Fig.~\ref{Fig:Toy}, we can see that a perfect photon-number resolving detector does not improve the QKD performance dramatically comparing to the threshold detector case.

\subsection{A few remarks} \label{Remark}
From the analysis of optimal $\mu$ in Appendix \ref{Optimalmu}, one can see that the key rate for the case without decoy states quadratically depends on the channel loss, $R=O(\eta^2)$, while for the case with decoy states, $R=O(\eta)$. This result is consistent with prior work in comparing the cases of coherent state QKD with and without decoy states \cite{Decoy_05}.

In the decoy state security proof \cite{Decoy_05}, the key assumption is that the decoy state and signal state should satisfy Eq.~\eqref{Passive:DecoyAss}.
This is guaranteed by the assumption that Eve cannot distinguish decoy and signal states. However, in the active decoy state method, Alice may introduce side information that can distinguish decoy and signal states when she actively prepares decoy and signal states. For example, an attenuator on Alice's side, used to prepare different intensities for signal and decoy states, may introduce different frequency shifts for signal and decoy states \cite{ZQMKQ_06}. In general, it is hard to verify the assumption that Eve cannot distinguish decoy and signal states in real active decoy state experiments.

In the passive decoy state scheme, decoy and signal states are passively determined by Alice's measurement outcome. Alice does not use an extra component (like in active decoy state method) to prepare decoy states. This reduces the possibility of side information leakage.
By passively choosing decoy states, Alice prepares same states on Bob's arm \footnote{Strictly speaking, there is one underlying assumption: the PDC source is single-mode.}. In fact, Alice can measure trigger signals after Bob finishes his measurements. Thus, to Eve's point of view, the states transmitted through the channel is independent of Alice's measurement results ($j$). Therefore, in principle, Eve cannot distinguish the decoy and signal states in the passive decoy state QKD.

This is the main advantage to use passive decoy state methods. Note that for coherent state QKD, one can only use active decoy state idea.

\section{Simulation} \label{Simulation}
In this section, we will compare the passive decoy state with a perfect number resolving detector and four QKD implementations with threshold detectors:  non-decoy, infinite decoy, weak active decoy and AYKI (passive decoy state).


We deduce experimental parameters from a recent PDC experiment \cite{PDC144_07},
which are listed in TABLE \ref{Tab:PDC144}. In the following simulations, we will use $q=1/2$ and $f(E_\mu)=1.22$ in Eq.~\eqref{Post:Keyj}. We notice that with the slightly modified experiment setup, a coherent state QKD with decoy states has been implemented \cite{PDC144_07}. Thus, it is reasonable to use this experiment setup to simulate the five QKD implementations.

\begin{table}[hbt]
\centering
\begin{tabular}{|c|c|c|c|c|c|c|c|c|c|c|} \hline
Frequency & Wavelength & $\eta_A$ & $\eta_{Bob}$ & $e_{d}$ & $Y_{0B}$ \\
\hline
249MHz & 710 nm & 14.5\% & 14.5\% & 1.5\% & $6.024\times 10^{-6}$ \\
\hline
\end{tabular}
\caption{List of parameters from 144 km PDC experiment \cite{PDC144_07}. Here $\eta_A$ and $\eta_{Bob}$ are the detection efficiencies in Alice and Bob's detection system, not including the optical channel loss. $e_d$ is the intrinsic detector error rate. $Y_{0B}$ is the background count rate of Bob's detection system (for example, if Bob has two detectors, then $Y_{0B}$ will be the sum of two detectors' background count rates). The transmission efficiency $\eta$ in Eq.~\eqref{Model:Yi} is given by $\eta_{Bob}$ plus the channel loss.} \label{Tab:PDC144}
\end{table}

In the simulation, for fair comparison, we always assume Bob uses the same detection setup (with threshold detectors).

\subsection{Without statistical fluctuations}
\begin{figure}[hbt]
\centering \resizebox{12cm}{!}{\includegraphics{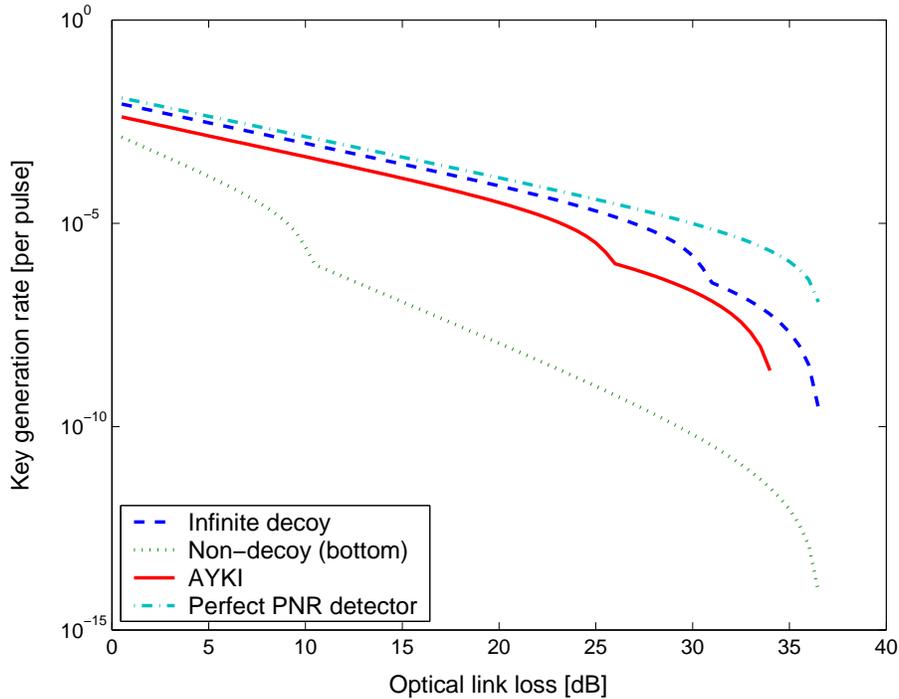}} \caption{Plot of the key
generation rate in terms of the optical loss, comparing four schemes without considering statistical fluctuations: non-decoy, infinite decoy, AYKI and the case with a perfect number resolving detector. Here we use $q=1/2$ and $f(E_\mu)=1.22$. We numerically optimize $\mu$ for each curve, see Appendix \ref{Optimalmu} for more discussions. Simulation parameters are listed in Table \ref{Tab:PDC144}.} \label{Fig:Toy}
\end{figure}

In the first simulation, we consider the case that Alice and Bob performs an infinitely long QKD (no statistical fluctuations). In this case the weak active decoy state protocol will approach the infinite decoy case \cite{Practical_05}. We assume that Alice is able to adjust $\mu$ (the brightness of the PDC source) in the regime of $[0,1]$ arbitrarily. In the simulation, we numerically optimize $\mu$ for each of the four implementation protocols: non-decoy, infinite decoy, AYKI and the case with a perfect number resolving detector. The simulation result is shown in FIG.~\ref{Fig:Toy}.

From FIG.~\ref{Fig:Toy}, we have the following remarks.
\begin{enumerate}
\item
In Appendix \ref{Optimalmu}, instead of numerically optimizing $\mu$ as done for Fig.~\eqref{Fig:Toy}, we qualitatively investigate the optimal $\mu$ for triggering PDC QKD with and without decoy states. The simulation result is consistent with the qualitative conclusion $R=O(\eta)$ for the case with decoy state and $R=O(\eta^2)$ for the case without decoy state.

\item
The space between the solid line and dashed line in FIG.~\ref{Fig:Toy} indicates the
room left for improvement by other decoy protocols with threshold detectors after
AYKI protocol is implemented. We can see that, in a large optical link loss regime,
the performances of AYKI and the infinite decoy are close. For instance, the AYKI
protocol yields around 50\% of the key rate of the infinite decoy state protocol
when the channel loss is within 20dB.

\item
By comparing AYKI and the case with a perfect photon-number resolving detector, we can see that even with a perfect photon-number resolving detector on Alice's side, the key rate is not improved dramatically in a large optical loss regime.



\item
The non-decoy protocol is better than AYKI in the regime close to maximal secure
distances. This is because we use the bounds of Eqs.~\eqref{Post:Y1T} and \eqref{Post:e1U} for AYKI curve. In reality, Alice and Bob can use the bound of Eq.~\eqref{Post:Q1e1nondecoy} directly in this regime.

\item
There is a bump in each curve. This is due to the fact that in the key generation
rate formula Eq.~\eqref{Post:Key01}, the non-triggered detection events have no
contribution to the final secure key after the bump.

\item
At the point of loss=0 dB, the key rates of four cases (from top to bottom) are $1.21\times10^{-2}$, $8.6\times10^{-3}$, $4.2\times10^{-3}$ and $1.3\times10^{-3}$.


\item
At the point of loss=0 dB, the numerical results for optimal $\mu$ for four cases (from top to bottom) are: 1, 0.52, 0.194, 0.0589. The optimal $\mu$ for the case with a perfect threshold detector is always 1, which is reasonable since $\mu=1$ maximizes the single photon state probability. In Appendix \ref{Optimalmu}, we show that the optimal $\mu$'s for the infinite decoy and AYKI case are relatively stable in a large optical loss regime. The optimal $\mu$ for the no decoy state case decreases with channel loss.

\item
We remark that the real $\mu$ used in the experiment \cite{PDC144_07} is $\mu=0.0265$. In general, it is experimentally hard to increase the brightness ($\mu$) of a PDC source.

\item
All of the four cases can tolerate similar optical losses.
\end{enumerate}


\subsection{With statistical fluctuations}
In a real experiment, the key length is always finite. Alice and Bob should consider statistical fluctuations. As pointed in Ref.~\cite{Practical_05}, statistical fluctuation analysis is a complicated problem in decoy state QKD.

Similar to the analysis in Ref.~\cite{Practical_05}, we assume a few conditions:
\begin{enumerate}
\item
Alice knows the exact value of average photon pair number $\mu$, which is a fixed number during key transmission.

\item
The distribution of photon number, Eq.~\eqref{Impl:PnL}, does not fluctuate.

\item
Assume the QKD transmission is part of an infinite length experiment.
\end{enumerate}

\begin{figure}[hbt]
\centering \resizebox{12cm}{!}{\includegraphics{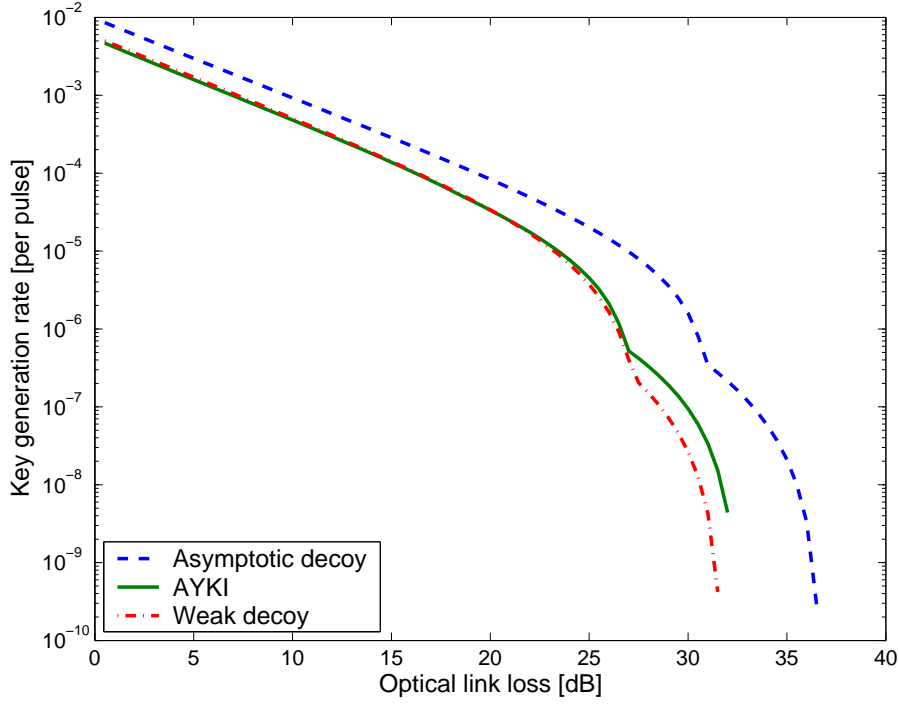}}
\caption{
Plot of the key generation rate in terms of the optical loss, comparing three cases with threshold detectors after considering statistical fluctuations: infinite decoy, weak active decoy and
AYKI. We numerically optimize $\mu$ for each curve. Here we use $q=1/2$ and $f(E_\mu)=1.22$. In the weak decoy state case, we assume Alice can randomly attenuate her PDC source intensity.  Simulation parameters are listed in Table \ref{Tab:PDC144}. The data size is $6\times10^9$ pumping laser pulses on Alice's side.
} \label{Fig:Fluctuation}
\end{figure}

Here we focus on the three cases with threshold detectors: infinite decoy, weak decoy and AYKI. We assume that the data size is $6\times10^9$ pulses of Alice's pumping laser. The simulation result is shown in FIG.~\ref{Fig:Fluctuation}. From the simulation result, we have the following observations.

\begin{enumerate}
\item
Similar to the case without fluctuation analysis, in a large optical
link loss regime, the performances of AYKI and the infinite decoy are close.

\item
At the point of loss=0 dB, the key rates of the three cases from top to bottom are $8.6\times10^{-3}$ (infinite), $5.0\times10^{-3}$ (weak) and $4.7\times10^{-3}$ (AYKI).

\item
The maximal tolerable secure optical losses for three cases are rather similar: 37dB (infinite), 32.5dB (AYKI), 32dB (weak).

\item
The AYKI protocol yields a higher key rate than weak decoy state protocol when the loss is greater than 16 dB. AYKI is less affected by statistical fluctuations than the weak decoy state because in AYKI, Alice does not need to sacrifice extra pulses for decoy states.


\end{enumerate}

In Section \ref{Remark}, we pointed out that from a practical security point of view, the passive decoy state method has an advantage over active decoy state methods. Also, AYKI method does not require any additional hardware change to implement decoy state, while in the weak decoy state case, Alice needs to add an attenuator to create decoy states. Now, from the simulation result, we can see that AYKI and weak active decoy state method yields similar QKD performance. Thus, our conclusion is that one should just use AYKI method instead of the weak decoy state method.

\section{Conclusion}
By investigating the optimal photon source intensity, we find that the triggering PDC QKD setup with decoy states is able to achieve a key rate that linearly depends on the channel transmittance, comparing to the quadratic dependence for the case without decoy states. Therefore, we expect the decoy state QKD to become a standard technique not only in the coherent state QKD, but also in QKD with triggering PDC sources.

On the practical side, we generalize the passive decoy state idea. The generalized passive decoy state idea can be applied to cases where either threshold detectors or photon number resolving detectors are used. The decoy protocol proposed by Adachi, Yamamoto, Koashi and Imoto (AYKI) can be treated as a special case of the generalized passive decoy state method. Comparing to the active (regular) decoy state methods, the passive one opens less possibility for Eve to distinguish decoy and signal states, which is a key underlying assumption in the security proof of decoy state QKD. From this sense, the passive decoy state method is more secure than the active decoy state methods in practice.

By simulating a recent PDC experiment, we compare various practical decoy state protocols with the infinite decoy protocol. We also compare the cases using threshold detectors and photon-number resolving detectors. Our simulation result shows that with the AYKI protocol, one can achieve a key generation rate that is close to the theoretical limit of infinite decoy protocol. Furthermore, our simulation result suggests that a photon-number resolving detector has little room to improve the QKD performance, comparing to the case using threshold detectors.

We also consider the statistical fluctuations. We compare infinite decoy protocol, weak active decoy state method and AYKI protocol. The simulation result shows that the weak active decoy state method and AYKI protocol yield very close QKD performance. In a large optical loss regime, the AYKI protocol can achieve a performance that is close to the infinite decoy case. Since the AYKI protocol requires no hardware changes for triggering PDC QKD, we conclude that AYKI method is a good protocol for triggering PDC QKD experiments.

Although our analysis is focused on the QKD with PDC sources, we emphasize that it can also be applied to other QKD setups with triggered single photon sources.


\section{Acknowledgments}
We thank C.-H.~F.~Fung, W.~Mauerer, A.~M.~Steinberg and G.~Weihs for enlightening discussions. In the simulation part, we thank H.~H\"{u}bel for confirming the parameters we deduced from their experiments. This work has been supported by CFI, CIFAR, CIPI, Connaught, CRC, MITACS, NSERC, OIT, PREA, QuantumWorks and the University of Toronto. X.~Ma gratefully acknowledges Chinese Government Award for Outstanding Self-financed Students Abroad and the Lachlan Gilchrist Fellowship. 

\begin{appendix}

\section{Optimal $\mu$} \label{Optimalmu}
The optimal $\mu$ for the coherent state QKD with and without decoy states has already been studied \cite{Practical_05}. Here instead of numerically optimizing $\mu$ as done for Fig.~\eqref{Fig:Toy}, we qualitatively investigate the optimal $\mu$ for the triggering PDC QKD with and without decoy states. Here we are interested in the case that Alice uses a threshold detector.

\subsection{Without decoy states}
Let us begin with the optimal $\mu$ of the case without decoy states. Here we will apply GLLP
\cite{GLLP_04} security analysis. As shown in Ref.~\cite{Low_06}, GLLP and L\"utkenhaus's
\cite{IndividualAttack_00} security analyses achieve similar performances for the coherent state
QKD. Intuitively, we should get a similar optimal $\mu$ as given in
Ref.~\cite{IndividualAttack_00}, $\mu\approx\eta/2$.

\begin{figure}[hbt]
\centering \resizebox{12cm}{!}{\includegraphics{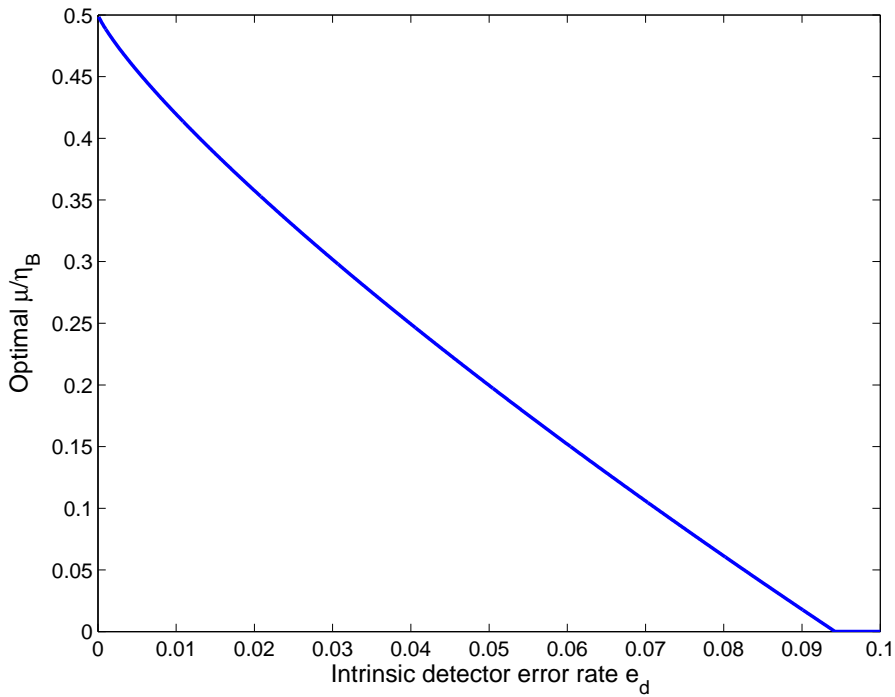}} \caption{Plot of the optimal $\mu$
in terms of $e_d$ for triggering PDC+non-decoy. Here we use $f(e_d)=1.22$ since the error rate is less 10\% \cite{BrassardSalvail_93}.} \label{Fig:optmunon}
\end{figure}

From Eq.~\eqref{Model:QEThreshold}, we can see that the gain $Q_{\mu,j}$ ($j=0,1$) is in the order of $\mu\eta$. To keep $Q_{1,0}$ or $Q_{1,1}$ in Eq.~\eqref{Post:Q1e1nondecoy} positive, $\mu$ should be in the order of $\eta$. By assuming $\mu$, $\eta$ and $Y_{0B}$ are small, we can simplify Eq.~\eqref{Model:QEThreshold}
\begin{equation}\label{Simulation:nondecoyApp}
\begin{aligned}
Q_{\mu,0}+Q_{\mu,1} &\approx \eta\mu \\
E_{\mu,0}\approx E_{\mu,0} &\approx e_d \\
Q^L_{1,0}+Q^L_{1,1} &\approx \eta\mu - \mu^2 \\
e^U_1 &\approx \frac{\eta e_d}{\eta-\mu} \\
\end{aligned}
\end{equation}
where $Q^L_{1,0}+Q^L_{1,1}$ is the lower bound of $Q_{1,0}+Q_{1,1}$ and $e^U_1$ is the upper bound of $e_1$ from Eq.~\eqref{Post:Q1e1nondecoy}. Since the error rates from triggered ($j=1$) and non-triggered ($j=0$) detection events are the same, the key generation rate given by Eq.~\eqref{Post:KeyTrig} can be simplified to
\begin{equation} \label{Simulation:KeySmartApp}
\begin{aligned}
R &\ge q \{-f(E_{\mu})Q_\mu H_2(E_{\mu})+Q_1[1-H_2(e_1)]+Q_0\} \\
  &\approx q \{-f(e_d)\eta\mu H_2(e_d)+(\eta\mu - \mu^2)[1-H_2(\frac{\eta e_d}{\eta-\mu})]\}
\end{aligned}
\end{equation}
By taking derivative of $R$, the optimal $\mu\equiv x\eta$ satisfies
\begin{equation} \label{Simulation:Keynondiffmu} 
\begin{aligned}
&-f(e_{d}) H_2(e_{d})+1-2x+e_d\log_2\frac{e_d}{1-x}+(1-e_d-2x)\log_2(1-\frac{e_d}{1-x})=0. \\
\end{aligned}
\end{equation}
Here if set $e_{d}=0$, then we get $x=1/2$ which is compatible with L\"ukenthaus' result
\cite{IndividualAttack_00}. We note that $x=1/2$ essentially maximize the probability of single photon source $Q^L_{1,0}+Q^L_{1,1}$ in Eq.~\eqref{Simulation:nondecoyApp}. More precisely, we can solve Eq.~\eqref{Simulation:Keynondiffmu} numerically, see FIG.~\ref{Fig:optmunon}.

From FIG.~\ref{Fig:optmunon}, we can see that the optimal $\mu$ for triggering PDC+non-decoy is
$\mu=O(\eta)$, which will lead the final key generation rate $R=O(\eta^2)$.

\subsection{With decoy states}
With decoy states, Alice and Bob can estimate $Q_1$ and $e_1$ better. Here we consider the
infinite decoy state case with threshold detectors. Under the assumption that $\eta$ and $Y_{0B}$ are small, we can simplify
Eqs.~\eqref{Model:QEThreshold} and \eqref{Model:qe1Threshold},
\begin{equation}\label{Simulation:AsymdecoyApp}
\begin{aligned}
Q_{\mu,0}+Q_{\mu,1} &\approx \eta\mu \\
E_{\mu,0} \approx E_{\mu,0} &\approx e_d \\
Q_{1,0}+Q_{1,1} &\approx \frac{\eta\mu}{(1+\mu)^2} \\
e_1 &\approx e_d \\
\end{aligned}
\end{equation}

\begin{figure}[hbt]
\centering \resizebox{12cm}{!}{\includegraphics{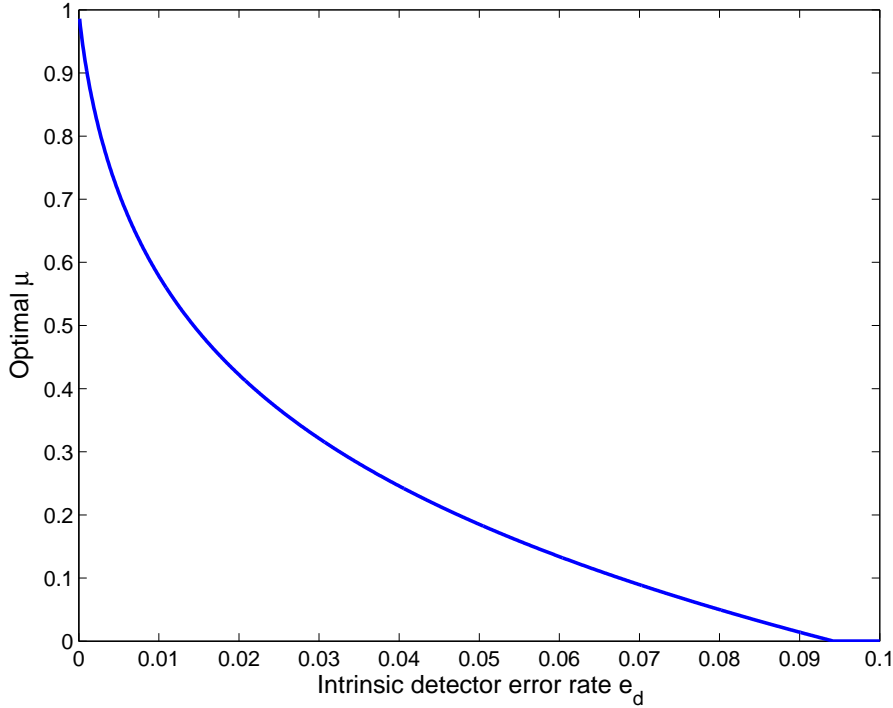}} \caption{Plot of the optimal $\mu$
in terms of $e_d$ for the triggering PDC+infinite decoy. Here we use$f(e_d)=1.22$.} \label{Fig:optmudec}
\end{figure}

With these approximations, the key generation rate given in Eq.~\eqref{Post:KeyTrig} can be
simplified to
\begin{equation} \label{Simulation:KeyTrigapp}
R \approx q \{-f(e_{d})\eta\mu H_2(e_{d})+ \frac{\eta\mu}{(1+\mu)^2}[1-H_2(e_{d})]\}.
\end{equation}
The optimal $\mu$ satisfies
\begin{equation} \label{Simulation:Keydecdiffmu}
\frac{1-\mu}{(1+\mu)^3}=\frac{f(e_{d})H_2(e_{d})}{1-H_2(e_{d})}
\end{equation}
Here if set $e_{d}=0$, then we get $\mu=1$ with which the probability to get a single photon state
is maximized. The numerical result of Eq.~\eqref{Simulation:Keydecdiffmu} is shown in
FIG.~\ref{Fig:optmudec}.


From FIG.~\ref{Fig:optmudec}, similar to the case coherent state QKD with decoy states
\cite{Practical_05}, one can see that the optimal $\mu$ is independent of channel loss $\eta$ for the infinite decoy state case with threshold detectors, i.e., $\mu=O(1)$, which will lead the final key generation rate $R=O(\eta)$.

\subsection{Numerical checking}
\begin{figure}[hbt]
\centering \resizebox{12cm}{!}{\includegraphics{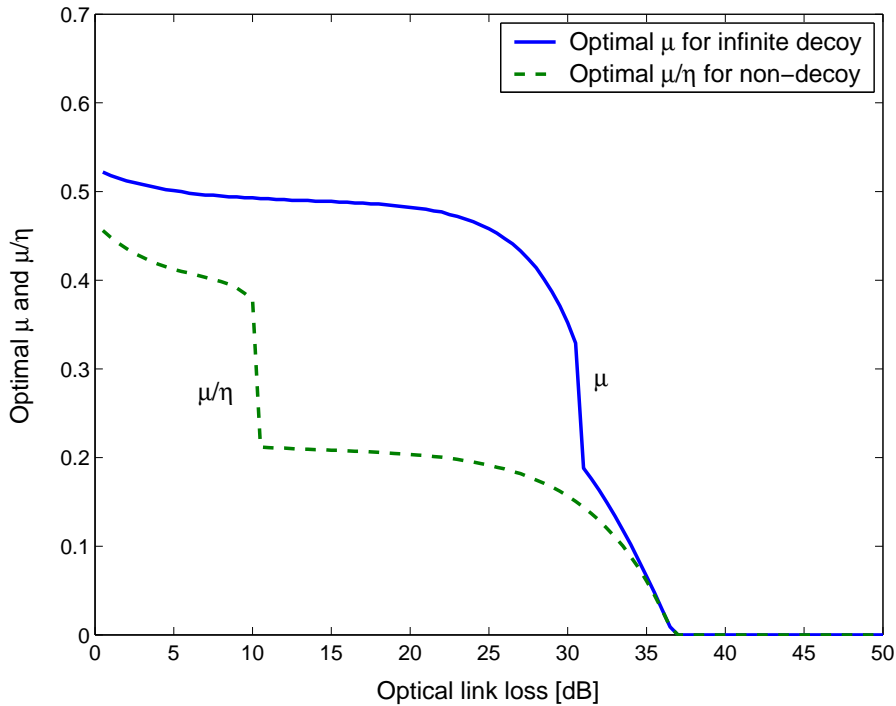}} \caption{Plot of the optimal $\mu$
in terms of optical loss for triggering PDC+non-decoy and triggering PDC+infinite-decoy. Here we use $q=1/2$ and $f(E_\mu)=1.22$. Simulation parameters are listed in Table \ref{Tab:PDC144}.} \label{Fig:munumer}
\end{figure}

Now we would like to numerically compare the optimal $\mu$ with and without decoy states by simulating a recent PDC experiment \cite{PDC144_07}, with parameters listed in Table \ref{Tab:PDC144}. In the simulation, we numerically optimize $\mu$ for the key rate given by Eq.~\eqref{Post:Key01} for the non-decoy and infinite decoy methods. For this particular setup, the optimal $\mu$ is shown in Figure \ref{Fig:munumer}.

From the figure we can see that the optimal $\mu$ for the non-decoy case is in the order of $\eta$, while the optimal $\mu$ for the infinite-decoy case is in the order of 1. This is consistent with the results of the analysis in the two previous subsections.

\end{appendix}

\bibliographystyle{apsrev}

\bibliography{Bibli}

\begin{thebibliography}{46}
\expandafter\ifx\csname natexlab\endcsname\relax\def\natexlab#1{#1}\fi
\expandafter\ifx\csname bibnamefont\endcsname\relax
  \def\bibnamefont#1{#1}\fi
\expandafter\ifx\csname bibfnamefont\endcsname\relax
  \def\bibfnamefont#1{#1}\fi
\expandafter\ifx\csname citenamefont\endcsname\relax
  \def\citenamefont#1{#1}\fi
\expandafter\ifx\csname url\endcsname\relax
  \def\url#1{\texttt{#1}}\fi
\expandafter\ifx\csname urlprefix\endcsname\relax\def\urlprefix{URL }\fi
\providecommand{\bibinfo}[2]{#2}
\providecommand{\eprint}[2][]{\url{#2}}

\bibitem[{\citenamefont{Bennett and Brassard}(1984)}]{BB_84}
\bibinfo{author}{\bibfnamefont{C.~H.} \bibnamefont{Bennett}} \bibnamefont{and}
  \bibinfo{author}{\bibfnamefont{G.}~\bibnamefont{Brassard}}, in
  \emph{\bibinfo{booktitle}{Proceedings of IEEE International Conference on
  Computers, Systems, and Signal Processing}} (\bibinfo{organization}{IEEE, New
  York}, \bibinfo{address}{Bangalore, India}, \bibinfo{year}{1984}), pp.
  \bibinfo{pages}{175--179}.

\bibitem[{\citenamefont{Ekert}(1991)}]{Ekert_91}
\bibinfo{author}{\bibfnamefont{A.~K.} \bibnamefont{Ekert}},
  \bibinfo{journal}{Phys.~Rev.~Lett.~} \textbf{\bibinfo{volume}{67}},
  \bibinfo{pages}{661} (\bibinfo{year}{1991}).

\bibitem[{\citenamefont{Mayers}(2001)}]{Mayers_01}
\bibinfo{author}{\bibfnamefont{D.}~\bibnamefont{Mayers}},
  \bibinfo{journal}{Journal of the ACM} \textbf{\bibinfo{volume}{48}},
  \bibinfo{pages}{351–406} (\bibinfo{year}{2001}).

\bibitem[{\citenamefont{Lo and Chau}(1999)}]{LoChauQKD_99}
\bibinfo{author}{\bibfnamefont{H.-K.} \bibnamefont{Lo}} \bibnamefont{and}
  \bibinfo{author}{\bibfnamefont{H.-F.} \bibnamefont{Chau}},
  \bibinfo{journal}{Science} \textbf{\bibinfo{volume}{283}},
  \bibinfo{pages}{2050} (\bibinfo{year}{1999}).

\bibitem[{\citenamefont{Shor and Preskill}(2000)}]{ShorPreskill_00}
\bibinfo{author}{\bibfnamefont{P.~W.} \bibnamefont{Shor}} \bibnamefont{and}
  \bibinfo{author}{\bibfnamefont{J.}~\bibnamefont{Preskill}},
  \bibinfo{journal}{Phys.~Rev.~Lett.~} \textbf{\bibinfo{volume}{85}},
  \bibinfo{pages}{441} (\bibinfo{year}{2000}).

\bibitem[{\citenamefont{Koashi}(2006{\natexlab{a}})}]{Koashi_Uncer_06}
\bibinfo{author}{\bibfnamefont{M.}~\bibnamefont{Koashi}}, \bibinfo{journal}{J.
  Phys. Conf. Ser.} \textbf{\bibinfo{volume}{36}}, \bibinfo{pages}{98}
  (\bibinfo{year}{2006}{\natexlab{a}}).

\bibitem[{\citenamefont{Mayers and Yao}(1998)}]{MayersYao_98}
\bibinfo{author}{\bibfnamefont{D.}~\bibnamefont{Mayers}} \bibnamefont{and}
  \bibinfo{author}{\bibfnamefont{A.}~\bibnamefont{Yao}}, in
  \emph{\bibinfo{booktitle}{FOCS, 39th Annual Symposium on Foundations of
  Computer Science}} (\bibinfo{publisher}{IEEE, Computer Society Press, Los
  Alamitos}, \bibinfo{year}{1998}), p. \bibinfo{pages}{503}.

\bibitem[{\citenamefont{L\"{u}tkenhaus}(2000)}]{IndividualAttack_00}
\bibinfo{author}{\bibfnamefont{N.}~\bibnamefont{L\"{u}tkenhaus}},
  \bibinfo{journal}{Phys.~Rev.~A} \textbf{\bibinfo{volume}{61}},
  \bibinfo{pages}{052304} (\bibinfo{year}{2000}).

\bibitem[{\citenamefont{Brassard et~al.}(2000)\citenamefont{Brassard,
  L\"{u}tkenhaus, Mor, and Sanders}}]{BLMS_00}
\bibinfo{author}{\bibfnamefont{G.}~\bibnamefont{Brassard}},
  \bibinfo{author}{\bibfnamefont{N.}~\bibnamefont{L\"{u}tkenhaus}},
  \bibinfo{author}{\bibfnamefont{T.}~\bibnamefont{Mor}}, \bibnamefont{and}
  \bibinfo{author}{\bibfnamefont{B.~C.} \bibnamefont{Sanders}},
  \bibinfo{journal}{Phys.~Rev.~Lett.~} \textbf{\bibinfo{volume}{85}},
  \bibinfo{pages}{1330} (\bibinfo{year}{2000}).

\bibitem[{\citenamefont{Inamori et~al.}(2007)\citenamefont{Inamori,
  L\"{u}tkenhaus, and Mayers}}]{ILM_07}
\bibinfo{author}{\bibfnamefont{H.}~\bibnamefont{Inamori}},
  \bibinfo{author}{\bibfnamefont{N.}~\bibnamefont{L\"{u}tkenhaus}},
  \bibnamefont{and} \bibinfo{author}{\bibfnamefont{D.}~\bibnamefont{Mayers}},
  \bibinfo{journal}{Eur. Phys. J. D} \textbf{\bibinfo{volume}{41}},
  \bibinfo{pages}{599} (\bibinfo{year}{2007}).

\bibitem[{\citenamefont{Koashi and Preskill}(2003)}]{KoashiPreskill_03}
\bibinfo{author}{\bibfnamefont{M.}~\bibnamefont{Koashi}} \bibnamefont{and}
  \bibinfo{author}{\bibfnamefont{J.}~\bibnamefont{Preskill}},
  \bibinfo{journal}{Phys.~Rev.~Lett.~} \textbf{\bibinfo{volume}{90}},
  \bibinfo{pages}{057902} (\bibinfo{year}{2003}).

\bibitem[{\citenamefont{Gottesman et~al.}(2004)\citenamefont{Gottesman, Lo,
  L\"utkenhaus, and Preskill}}]{GLLP_04}
\bibinfo{author}{\bibfnamefont{D.}~\bibnamefont{Gottesman}},
  \bibinfo{author}{\bibfnamefont{H.-K.} \bibnamefont{Lo}},
  \bibinfo{author}{\bibfnamefont{N.}~\bibnamefont{L\"utkenhaus}},
  \bibnamefont{and} \bibinfo{author}{\bibfnamefont{J.}~\bibnamefont{Preskill}},
  \bibinfo{journal}{Quantum Information and Computation}
  \textbf{\bibinfo{volume}{4}}, \bibinfo{pages}{325} (\bibinfo{year}{2004}).

\bibitem[{\citenamefont{Bennett et~al.}(1992)\citenamefont{Bennett, Bessette,
  Brassard, Salvail, and Smolin}}]{BBBSS_92}
\bibinfo{author}{\bibfnamefont{C.~H.} \bibnamefont{Bennett}},
  \bibinfo{author}{\bibfnamefont{F.}~\bibnamefont{Bessette}},
  \bibinfo{author}{\bibfnamefont{G.}~\bibnamefont{Brassard}},
  \bibinfo{author}{\bibfnamefont{L.}~\bibnamefont{Salvail}}, \bibnamefont{and}
  \bibinfo{author}{\bibfnamefont{J.~A.} \bibnamefont{Smolin}},
  \bibinfo{journal}{Journal of Cryptology} \textbf{\bibinfo{volume}{5}},
  \bibinfo{pages}{3} (\bibinfo{year}{1992}).

\bibitem[{\citenamefont{Huttner et~al.}(1995)\citenamefont{Huttner, Imoto,
  Gisin, and Mor}}]{HIGM_95}
\bibinfo{author}{\bibfnamefont{B.}~\bibnamefont{Huttner}},
  \bibinfo{author}{\bibfnamefont{N.}~\bibnamefont{Imoto}},
  \bibinfo{author}{\bibfnamefont{N.}~\bibnamefont{Gisin}}, \bibnamefont{and}
  \bibinfo{author}{\bibfnamefont{T.}~\bibnamefont{Mor}},
  \bibinfo{journal}{Phys.~Rev.~A} \textbf{\bibinfo{volume}{51}},
  \bibinfo{pages}{1863} (\bibinfo{year}{1995}).

\bibitem[{\citenamefont{L\"{u}tkenhaus and Jahma}(2002)}]{LutkenhausJahma_02}
\bibinfo{author}{\bibfnamefont{N.}~\bibnamefont{L\"{u}tkenhaus}}
  \bibnamefont{and} \bibinfo{author}{\bibfnamefont{M.}~\bibnamefont{Jahma}},
  \bibinfo{journal}{New Journal of Physics} \textbf{\bibinfo{volume}{4}},
  \bibinfo{pages}{44.1} (\bibinfo{year}{2002}).

\bibitem[{\citenamefont{Hwang}(2003)}]{Hwang_03}
\bibinfo{author}{\bibfnamefont{W.-Y.} \bibnamefont{Hwang}},
  \bibinfo{journal}{Phys.~Rev.~Lett.~} \textbf{\bibinfo{volume}{91}},
  \bibinfo{pages}{057901} (\bibinfo{year}{2003}).

\bibitem[{\citenamefont{Lo}(2004)}]{LoDecoy_03}
\bibinfo{author}{\bibfnamefont{H.-K.} \bibnamefont{Lo}}, in
  \emph{\bibinfo{booktitle}{Proc.~of IEEE ISIT}} (\bibinfo{publisher}{IEEE},
  \bibinfo{year}{2004}), p. \bibinfo{pages}{137}.

\bibitem[{\citenamefont{Ma}(2004)}]{MasterReport}
\bibinfo{author}{\bibfnamefont{X.}~\bibnamefont{Ma}}, \bibinfo{journal}{arXiv:
  quant-ph/0503057}  (\bibinfo{year}{2004}).

\bibitem[{\citenamefont{Lo et~al.}(2005{\natexlab{a}})\citenamefont{Lo, Ma, and
  Chen}}]{Decoy_05}
\bibinfo{author}{\bibfnamefont{H.-K.} \bibnamefont{Lo}},
  \bibinfo{author}{\bibfnamefont{X.}~\bibnamefont{Ma}}, \bibnamefont{and}
  \bibinfo{author}{\bibfnamefont{K.}~\bibnamefont{Chen}},
  \bibinfo{journal}{Phys.~Rev.~Lett.~} \textbf{\bibinfo{volume}{94}},
  \bibinfo{pages}{230504} (\bibinfo{year}{2005}{\natexlab{a}}).

\bibitem[{\citenamefont{Ma et~al.}(2005)\citenamefont{Ma, Qi, Zhao, and
  Lo}}]{Practical_05}
\bibinfo{author}{\bibfnamefont{X.}~\bibnamefont{Ma}},
  \bibinfo{author}{\bibfnamefont{B.}~\bibnamefont{Qi}},
  \bibinfo{author}{\bibfnamefont{Y.}~\bibnamefont{Zhao}}, \bibnamefont{and}
  \bibinfo{author}{\bibfnamefont{H.-K.} \bibnamefont{Lo}},
  \bibinfo{journal}{Phys.~Rev.~A} \textbf{\bibinfo{volume}{72}},
  \bibinfo{pages}{012326} (\bibinfo{year}{2005}).

\bibitem[{\citenamefont{Harrington et~al.}(2005)\citenamefont{Harrington,
  Ettinger, Hughes, and Nordholt}}]{HEHN_05}
\bibinfo{author}{\bibfnamefont{J.~W.} \bibnamefont{Harrington}},
  \bibinfo{author}{\bibfnamefont{J.~M.} \bibnamefont{Ettinger}},
  \bibinfo{author}{\bibfnamefont{R.~J.} \bibnamefont{Hughes}},
  \bibnamefont{and} \bibinfo{author}{\bibfnamefont{J.~E.}
  \bibnamefont{Nordholt}}, \bibinfo{journal}{ArXiv.org:quant-ph/0503002}
  (\bibinfo{year}{2005}).

\bibitem[{\citenamefont{Wang}(2005{\natexlab{a}})}]{Wang_05}
\bibinfo{author}{\bibfnamefont{X.-B.} \bibnamefont{Wang}},
  \bibinfo{journal}{Phys.~Rev.~Lett.~} \textbf{\bibinfo{volume}{94}},
  \bibinfo{pages}{230503} (\bibinfo{year}{2005}{\natexlab{a}}).

\bibitem[{\citenamefont{Wang}(2005{\natexlab{b}})}]{Wang2_05}
\bibinfo{author}{\bibfnamefont{X.-B.} \bibnamefont{Wang}},
  \bibinfo{journal}{Phys.~Rev.~A} \textbf{\bibinfo{volume}{72}},
  \bibinfo{pages}{012322} (\bibinfo{year}{2005}{\natexlab{b}}).

\bibitem[{\citenamefont{Zhao et~al.}(2006{\natexlab{a}})\citenamefont{Zhao, Qi,
  Ma, Lo, and Qian}}]{ZQMKQ_06}
\bibinfo{author}{\bibfnamefont{Y.}~\bibnamefont{Zhao}},
  \bibinfo{author}{\bibfnamefont{B.}~\bibnamefont{Qi}},
  \bibinfo{author}{\bibfnamefont{X.}~\bibnamefont{Ma}},
  \bibinfo{author}{\bibfnamefont{H.-K.} \bibnamefont{Lo}}, \bibnamefont{and}
  \bibinfo{author}{\bibfnamefont{L.}~\bibnamefont{Qian}},
  \bibinfo{journal}{Phys.~Rev.~Lett.~} \textbf{\bibinfo{volume}{96}},
  \bibinfo{pages}{070502} (\bibinfo{year}{2006}{\natexlab{a}}).

\bibitem[{\citenamefont{Zhao et~al.}(2006{\natexlab{b}})\citenamefont{Zhao, Qi,
  Ma, Lo, and Qian}}]{ZQMKQ60km_06}
\bibinfo{author}{\bibfnamefont{Y.}~\bibnamefont{Zhao}},
  \bibinfo{author}{\bibfnamefont{B.}~\bibnamefont{Qi}},
  \bibinfo{author}{\bibfnamefont{X.}~\bibnamefont{Ma}},
  \bibinfo{author}{\bibfnamefont{H.-K.} \bibnamefont{Lo}}, \bibnamefont{and}
  \bibinfo{author}{\bibfnamefont{L.}~\bibnamefont{Qian}}, in
  \emph{\bibinfo{booktitle}{Proc.~of IEEE ISIT}} (\bibinfo{publisher}{IEEE},
  \bibinfo{year}{2006}{\natexlab{b}}), p. \bibinfo{pages}{2094}.

\bibitem[{\citenamefont{Rosenberg et~al.}(2007)\citenamefont{Rosenberg,
  Harrington, Rice, Hiskett, Peterson, Hughes, Lita, Nam, and
  Nordholt}}]{LosAlamosDecoy_07}
\bibinfo{author}{\bibfnamefont{D.}~\bibnamefont{Rosenberg}},
  \bibinfo{author}{\bibfnamefont{J.~W.} \bibnamefont{Harrington}},
  \bibinfo{author}{\bibfnamefont{P.~R.} \bibnamefont{Rice}},
  \bibinfo{author}{\bibfnamefont{P.~A.} \bibnamefont{Hiskett}},
  \bibinfo{author}{\bibfnamefont{C.~G.} \bibnamefont{Peterson}},
  \bibinfo{author}{\bibfnamefont{R.~J.} \bibnamefont{Hughes}},
  \bibinfo{author}{\bibfnamefont{A.~E.} \bibnamefont{Lita}},
  \bibinfo{author}{\bibfnamefont{S.~W.} \bibnamefont{Nam}}, \bibnamefont{and}
  \bibinfo{author}{\bibfnamefont{J.~E.} \bibnamefont{Nordholt}},
  \bibinfo{journal}{Phys.~Rev.~Lett.~} \textbf{\bibinfo{volume}{98}},
  \bibinfo{pages}{010503} (\bibinfo{year}{2007}).

\bibitem[{\citenamefont{Schmitt-Manderbach
  et~al.}(2007)\citenamefont{Schmitt-Manderbach, Weier, F\"urst, Ursin,
  Tiefenbacher, Scheidl, Perdigues, Sodnik, Kurtsiefer, Rarity
  et~al.}}]{Zeilinger_Decoy_07}
\bibinfo{author}{\bibfnamefont{T.}~\bibnamefont{Schmitt-Manderbach}},
  \bibinfo{author}{\bibfnamefont{H.}~\bibnamefont{Weier}},
  \bibinfo{author}{\bibfnamefont{M.}~\bibnamefont{F\"urst}},
  \bibinfo{author}{\bibfnamefont{R.}~\bibnamefont{Ursin}},
  \bibinfo{author}{\bibfnamefont{F.}~\bibnamefont{Tiefenbacher}},
  \bibinfo{author}{\bibfnamefont{T.}~\bibnamefont{Scheidl}},
  \bibinfo{author}{\bibfnamefont{J.}~\bibnamefont{Perdigues}},
  \bibinfo{author}{\bibfnamefont{Z.}~\bibnamefont{Sodnik}},
  \bibinfo{author}{\bibfnamefont{C.}~\bibnamefont{Kurtsiefer}},
  \bibinfo{author}{\bibfnamefont{J.~G.} \bibnamefont{Rarity}},
  \bibnamefont{et~al.}, \bibinfo{journal}{Phys.~Rev.~Lett.~}
  \textbf{\bibinfo{volume}{98}}, \bibinfo{pages}{010504}
  (\bibinfo{year}{2007}).

\bibitem[{\citenamefont{Peng et~al.}(2007)\citenamefont{Peng, Zhang, Yang, Gao,
  Ma, Yin, Zeng, Yang, Wang, and Pan}}]{PanDecoy_07}
\bibinfo{author}{\bibfnamefont{C.-Z.} \bibnamefont{Peng}},
  \bibinfo{author}{\bibfnamefont{J.}~\bibnamefont{Zhang}},
  \bibinfo{author}{\bibfnamefont{D.}~\bibnamefont{Yang}},
  \bibinfo{author}{\bibfnamefont{W.-B.} \bibnamefont{Gao}},
  \bibinfo{author}{\bibfnamefont{H.-X.} \bibnamefont{Ma}},
  \bibinfo{author}{\bibfnamefont{H.}~\bibnamefont{Yin}},
  \bibinfo{author}{\bibfnamefont{H.-P.} \bibnamefont{Zeng}},
  \bibinfo{author}{\bibfnamefont{T.}~\bibnamefont{Yang}},
  \bibinfo{author}{\bibfnamefont{X.-B.} \bibnamefont{Wang}}, \bibnamefont{and}
  \bibinfo{author}{\bibfnamefont{J.-W.} \bibnamefont{Pan}},
  \bibinfo{journal}{Phys.~Rev.~Lett.~} \textbf{\bibinfo{volume}{98}},
  \bibinfo{pages}{010505} (\bibinfo{year}{2007}).

\bibitem[{\citenamefont{Yuan et~al.}(2007)\citenamefont{Yuan, Sharpe, and
  Shields}}]{YSS_Decoy_07}
\bibinfo{author}{\bibfnamefont{Z.~L.} \bibnamefont{Yuan}},
  \bibinfo{author}{\bibfnamefont{A.~W.} \bibnamefont{Sharpe}},
  \bibnamefont{and} \bibinfo{author}{\bibfnamefont{A.~J.}
  \bibnamefont{Shields}}, \bibinfo{journal}{Appl.~Phys.~Lett.}
  \textbf{\bibinfo{volume}{90}}, \bibinfo{pages}{011118}
  (\bibinfo{year}{2007}).

\bibitem[{\citenamefont{Koashi}(2004)}]{Koashi_04}
\bibinfo{author}{\bibfnamefont{M.}~\bibnamefont{Koashi}},
  \bibinfo{journal}{Phys.~Rev.~Lett.~} \textbf{\bibinfo{volume}{93}},
  \bibinfo{pages}{120501} (\bibinfo{year}{2004}).

\bibitem[{\citenamefont{Tamaki et~al.}(2006)\citenamefont{Tamaki, L\"utkenhaus,
  Koashi, and Batuwantudawe}}]{TLMB_Strong06}
\bibinfo{author}{\bibfnamefont{K.}~\bibnamefont{Tamaki}},
  \bibinfo{author}{\bibfnamefont{N.}~\bibnamefont{L\"utkenhaus}},
  \bibinfo{author}{\bibfnamefont{M.}~\bibnamefont{Koashi}}, \bibnamefont{and}
  \bibinfo{author}{\bibfnamefont{J.}~\bibnamefont{Batuwantudawe}},
  \bibinfo{journal}{arXiv:quant-ph/0607082}  (\bibinfo{year}{2006}).

\bibitem[{\citenamefont{Inoue et~al.}(2002)\citenamefont{Inoue, Waks, and
  Yamamoto}}]{IWY_DPS02}
\bibinfo{author}{\bibfnamefont{K.}~\bibnamefont{Inoue}},
  \bibinfo{author}{\bibfnamefont{E.}~\bibnamefont{Waks}}, \bibnamefont{and}
  \bibinfo{author}{\bibfnamefont{Y.}~\bibnamefont{Yamamoto}},
  \bibinfo{journal}{Phys.~Rev.~Lett.~} \textbf{\bibinfo{volume}{89}},
  \bibinfo{pages}{037902} (\bibinfo{year}{2002}).

\bibitem[{\citenamefont{Ma et~al.}(2007)\citenamefont{Ma, Fung, and
  Lo}}]{EntanglementPDC_07}
\bibinfo{author}{\bibfnamefont{X.}~\bibnamefont{Ma}},
  \bibinfo{author}{\bibfnamefont{C.-H.~F.} \bibnamefont{Fung}},
  \bibnamefont{and} \bibinfo{author}{\bibfnamefont{H.-K.} \bibnamefont{Lo}},
  \bibinfo{journal}{Phys.~Rev.~A} \textbf{\bibinfo{volume}{76}},
  \bibinfo{pages}{012307} (\bibinfo{year}{2007}).

\bibitem[{\citenamefont{Mauerer and Silberhorn}(2007)}]{MauererSilberhorn_07}
\bibinfo{author}{\bibfnamefont{W.}~\bibnamefont{Mauerer}} \bibnamefont{and}
  \bibinfo{author}{\bibfnamefont{C.}~\bibnamefont{Silberhorn}},
  \bibinfo{journal}{Phys.~Rev.~A} \textbf{\bibinfo{volume}{75}},
  \bibinfo{pages}{050305(R)} (\bibinfo{year}{2007}).

\bibitem[{\citenamefont{Adachi et~al.}(2007)\citenamefont{Adachi, Yamamoto,
  Koashi, and Imoto}}]{AYKI_07}
\bibinfo{author}{\bibfnamefont{Y.}~\bibnamefont{Adachi}},
  \bibinfo{author}{\bibfnamefont{T.}~\bibnamefont{Yamamoto}},
  \bibinfo{author}{\bibfnamefont{M.}~\bibnamefont{Koashi}}, \bibnamefont{and}
  \bibinfo{author}{\bibfnamefont{N.}~\bibnamefont{Imoto}},
  \bibinfo{journal}{Phys.~Rev.~Lett.~} \textbf{\bibinfo{volume}{99}},
  \bibinfo{pages}{180503} (\bibinfo{year}{2007}).

\bibitem[{\citenamefont{Wang et~al.}(2007{\natexlab{a}})\citenamefont{Wang,
  Wang, and Guo}}]{WWG_07}
\bibinfo{author}{\bibfnamefont{Q.}~\bibnamefont{Wang}},
  \bibinfo{author}{\bibfnamefont{X.-B.} \bibnamefont{Wang}}, \bibnamefont{and}
  \bibinfo{author}{\bibfnamefont{G.-C.} \bibnamefont{Guo}},
  \bibinfo{journal}{Phys.~Rev.~A} \textbf{\bibinfo{volume}{75}},
  \bibinfo{pages}{012312} (\bibinfo{year}{2007}{\natexlab{a}}).

\bibitem[{\citenamefont{Wang et~al.}(2007{\natexlab{b}})\citenamefont{Wang,
  Wang, Bj\"ork, and Karlsson}}]{WWBK_PDC_07}
\bibinfo{author}{\bibfnamefont{Q.}~\bibnamefont{Wang}},
  \bibinfo{author}{\bibfnamefont{X.-B.} \bibnamefont{Wang}},
  \bibinfo{author}{\bibfnamefont{G.}~\bibnamefont{Bj\"ork}}, \bibnamefont{and}
  \bibinfo{author}{\bibfnamefont{A.}~\bibnamefont{Karlsson}},
  \bibinfo{journal}{Europhysics Letters} \textbf{\bibinfo{volume}{79}},
  \bibinfo{pages}{40001} (\bibinfo{year}{2007}{\natexlab{b}}).

\bibitem[{\citenamefont{Ursin et~al.}(2007)\citenamefont{Ursin, Tiefenbacher,
  Schmitt-Manderbach, Weier, Scheidl, Lindenthal, Blauensteiner, Jennewein,
  Perdigues, Trojek et~al.}}]{PDC144_07}
\bibinfo{author}{\bibfnamefont{R.}~\bibnamefont{Ursin}},
  \bibinfo{author}{\bibfnamefont{F.}~\bibnamefont{Tiefenbacher}},
  \bibinfo{author}{\bibfnamefont{T.}~\bibnamefont{Schmitt-Manderbach}},
  \bibinfo{author}{\bibfnamefont{H.}~\bibnamefont{Weier}},
  \bibinfo{author}{\bibfnamefont{T.}~\bibnamefont{Scheidl}},
  \bibinfo{author}{\bibfnamefont{M.}~\bibnamefont{Lindenthal}},
  \bibinfo{author}{\bibfnamefont{B.}~\bibnamefont{Blauensteiner}},
  \bibinfo{author}{\bibfnamefont{T.}~\bibnamefont{Jennewein}},
  \bibinfo{author}{\bibfnamefont{J.}~\bibnamefont{Perdigues}},
  \bibinfo{author}{\bibfnamefont{P.}~\bibnamefont{Trojek}},
  \bibnamefont{et~al.}, \bibinfo{journal}{Nature Physics}
  \textbf{\bibinfo{volume}{3}}, \bibinfo{pages}{481} (\bibinfo{year}{2007}).

\bibitem[{\citenamefont{Walls and Milburn}(1994)}]{WallsMilburn_94}
\bibinfo{author}{\bibfnamefont{D.~F.} \bibnamefont{Walls}} \bibnamefont{and}
  \bibinfo{author}{\bibfnamefont{G.~J.} \bibnamefont{Milburn}},
  \emph{\bibinfo{title}{Quantum Optics}} (\bibinfo{publisher}{Springer,
  Berlin}, \bibinfo{year}{1994}).

\bibitem[{\citenamefont{Scarani et~al.}(2004)\citenamefont{Scarani, A.~Acin,
  and Gisin}}]{SARG_04}
\bibinfo{author}{\bibfnamefont{V.}~\bibnamefont{Scarani}},
  \bibinfo{author}{\bibfnamefont{G.~R.} \bibnamefont{A.~Acin}},
  \bibnamefont{and} \bibinfo{author}{\bibfnamefont{N.}~\bibnamefont{Gisin}},
  \bibinfo{journal}{Phys.~Rev.~Lett.~} \textbf{\bibinfo{volume}{92}},
  \bibinfo{pages}{057901} (\bibinfo{year}{2004}).

\bibitem[{\citenamefont{Lo}(2005)}]{Vacuum_05}
\bibinfo{author}{\bibfnamefont{H.-K.} \bibnamefont{Lo}},
  \bibinfo{journal}{Quantum Information and Computation}
  \textbf{\bibinfo{volume}{5}}, \bibinfo{pages}{413} (\bibinfo{year}{2005}).

\bibitem[{\citenamefont{Koashi}(2006{\natexlab{b}})}]{Koashi_NewModel_06}
\bibinfo{author}{\bibfnamefont{M.}~\bibnamefont{Koashi}},
  \bibinfo{journal}{arXiv:quant-ph/0609180}
  (\bibinfo{year}{2006}{\natexlab{b}}).

\bibitem[{\citenamefont{Lo et~al.}(2005{\natexlab{b}})\citenamefont{Lo, Chau,
  and Ardehali}}]{EffBB84_05}
\bibinfo{author}{\bibfnamefont{H.-K.} \bibnamefont{Lo}},
  \bibinfo{author}{\bibfnamefont{H.-F.} \bibnamefont{Chau}}, \bibnamefont{and}
  \bibinfo{author}{\bibfnamefont{M.}~\bibnamefont{Ardehali}},
  \bibinfo{journal}{Journal of Cryptology} \textbf{\bibinfo{volume}{18}},
  \bibinfo{pages}{133} (\bibinfo{year}{2005}{\natexlab{b}}).

\bibitem[{\citenamefont{Brassard and Salvail}(1993)}]{BrassardSalvail_93}
\bibinfo{author}{\bibfnamefont{G.}~\bibnamefont{Brassard}} \bibnamefont{and}
  \bibinfo{author}{\bibfnamefont{L.}~\bibnamefont{Salvail}}, in
  \emph{\bibinfo{booktitle}{Advances in Cryptology EUROCRYPT '93}}, edited by
  \bibinfo{editor}{\bibfnamefont{G.}~\bibnamefont{Goos}} \bibnamefont{and}
  \bibinfo{editor}{\bibfnamefont{J.}~\bibnamefont{Hartmanis}}
  (\bibinfo{publisher}{Springer-Verlag, Berlin}, \bibinfo{year}{1993}).

\bibitem[{\citenamefont{Ma et~al.}(2006)\citenamefont{Ma, Fung, Dupuis, Chen,
  Tamaki, and Lo}}]{TwoWay_06}
\bibinfo{author}{\bibfnamefont{X.}~\bibnamefont{Ma}},
  \bibinfo{author}{\bibfnamefont{C.-H.~F.} \bibnamefont{Fung}},
  \bibinfo{author}{\bibfnamefont{F.}~\bibnamefont{Dupuis}},
  \bibinfo{author}{\bibfnamefont{K.}~\bibnamefont{Chen}},
  \bibinfo{author}{\bibfnamefont{K.}~\bibnamefont{Tamaki}}, \bibnamefont{and}
  \bibinfo{author}{\bibfnamefont{H.-K.} \bibnamefont{Lo}},
  \bibinfo{journal}{Phys.~Rev.~A} \textbf{\bibinfo{volume}{74}},
  \bibinfo{pages}{032330} (\bibinfo{year}{2006}).

\bibitem[{\citenamefont{Ma}(2006)}]{Low_06}
\bibinfo{author}{\bibfnamefont{X.}~\bibnamefont{Ma}},
  \bibinfo{journal}{Phys.~Rev.~A} \textbf{\bibinfo{volume}{74}},
  \bibinfo{pages}{052325} (\bibinfo{year}{2006}).

\end{thebibliography}


\end{document}